\documentclass[aps,prd,showpacs,preprint,nofootinbib,onecolumn]{revtex4-1}
\usepackage{graphicx} % Required for inserting images
\usepackage{multirow}
\usepackage{rotating}
\usepackage{float}
\usepackage{amsmath}
\usepackage{url}
\usepackage[utf8]{inputenc}

\begin{document}

\title{Pseudo-Newtonian potential for accretion disks in a modified gravity spacetime around the black hole and underlying properties}

\author{Sriraj Chandra}
\email{srirajc@iisc.ac.in}
\affiliation{Department of Physics, Indian Institute of Science, Bangalore 560012}

\author{Banibrata Mukhopadhyay}
\email{bm@iisc.ac.in}
\affiliation{Department of Physics, Indian Institute of Science, Bangalore 560012}

\date{\today}

\begin{abstract}

We construct a pseudo-Newtonian potential (PNP) corresponding to a rotating black hole solution in a modified gravity (MGR) framework using a metric-based prescription. The motivation is to enable realistic accretion disk studies in MGR, where full relativistic MHD simulations remain computationally prohibitive. Effective potentials and the underlying Newtonian-like forces are derived for both massive and massless particles in the equatorial plane, relevant for disk dynamics. The reliability of the PNP is tested by comparing key orbital properties -- marginally stable, marginally bound, photon orbits and energies at marginally stable orbit radii -- with exact MGR predictions. The PNP reproduces the marginally stable and photon orbits exactly, while marginally bound orbits and specific energies deviate by less than about 7–10\%. The influence of the MGR parameter on particle dynamics and effective potentials is analyzed, revealing non-trivial departures from simple Newtonian intuition. The study demonstrates that the proposed PNP accurately captures essential spatial properties of MGR spacetime and provides an efficient, physically consistent tool for investigating accretion phenomena and strong-gravity astrophysics beyond general relativity.
    
\end{abstract}

\maketitle

\section{Introduction}
\label{intro}

Black holes provide the most extreme natural laboratories for testing gravity in the strong-field regime. As black holes cannot be seen directly, the underlying accretion disk forming around them is a very important astrophysical object. Accretion flows around black holes are intrinsically relativistic and become supersonic in the innermost regions, unlike flows around less compact objects like neutron stars and white dwarf. Depending on their temperature, composition, and magnetic field content, these flows are best described within the framework of general relativistic magnetohydrodynamics (GRMHD) (e.g., \cite{Gammie2003, Mckinney2014, Porth2019}). Such accretion processes power active galactic nuclei, X-ray binaries, ultraluminous X-ray sources, and relativistic jets, and are now directly probed by high-resolution observations such as those of the Event Horizon Telescope \cite{EHT2019, EHT2022}.

General relativity (GR) has been remarkably successful in explaining gravitational phenomena over a wide range of scales. Precision tests in the Solar System \cite{Will2014}, binary pulsars \cite{Kramer2006}, and gravitational-wave (GW) detections from compact binaries \cite{abbott2016, abbott2019} have provided strong support for Einstein’s theory. Nevertheless, most of these tests probe either weak or moderately strong gravitational fields. The truly strong-field regime near black hole horizons remains comparatively unexplored observationally, leaving open the possibility that GR may require modification in such extreme environments \cite{berti2015, yunes2013}.

Motivated by this, a wide class of modified gravity (MGR) theories has been proposed, including scalar–tensor theories: $f(R)$-gravity, Gauss–Bonnet models, and higher-curvature extensions \cite{clifton2012, nojiri2017, capozziello2011, sotiriou2010}. These theories are often constructed in such a way that they recover GR in appropriate limits, while allowing deviations in the strong-field regime. These MGR theories have found various applications: from cosmology to compact object physics, influencing the structure of neutron stars, white dwarfs, and black holes \cite{psaltis2008, doneva2018, berti2018}.

In the stellar context, MGR has been shown to alter mass–radius relations of white dwarfs and neutron stars, potentially explaining observational anomalies in Type Ia supernova progenitors, massive white dwarfs and neutron stars \cite{Kalita-MukhWD, astashenok2013, yazadjiev2014}. Our group has demonstrated that MGR can unify sub- and super-Chandrasekhar limiting mass of white dwarfs within a single theoretical framework \cite{Kalita-MukhWD, UDas-Mukh}, and the deviations from GR may leave detectable imprints in GW observations \cite{Kalita-MukhApJMountain}.

In the black hole sector, nonrotating and rotating solutions in MGR exhibit systematic deviations from Schwarzschild and Kerr geometries, respectively, including the shifts in the radii of event horizon ($r_h$), marginally stable circular orbit ($r_{ms}$), marginally bound orbit ($r_{mb}$), photon orbit ($r_{ph}$) and energy at $r_{ms}$ ($E_{ms}$) \cite{surajitModBH,ADas-Mukh}. Such deviations can influence accretion disk structure, radiative efficiency, spectral properties, and quasi-periodic oscillations \cite{kato2001, abramowicz2013, bambi2017}. Consequently, accretion disks provide a powerful observational probe of strong-field gravity and potential deviations from GR.

From a computational standpoint, GRMHD simulations currently represent the most realistic approach to modeling black hole accretion \cite{Gammie2003, mckinney2012, narayan2012, EHT2019}. However, these simulations are formulated strictly within GR. Fully relativistic MHD simulations in MGR are not yet available, primarily due to the mathematical and numerical complexity of MGR field equations and also uncertainly of MGR theories. This severely limits our ability to explore MGR effects in realistic accretion environments using direct numerical approaches.

The pseudo-Newtonian potentials (PNPs) have long been used as effective tools for capturing essential general relativistic features within Newtonian hydrodynamics. The Paczyński–Wiita potential for Schwarzschild black holes is a classic example \cite{PW80}, and numerous generalizations have been developed for rotating black holes and frame-dragging effects \cite{Chakrabarti-Khanna, mondal-chakrabarti2006, ArtemovaPNP,Ghosh-Mukh07, semerak1999}. Hence, a PNP will be very useful to overcome the limitation to perform accretion disk simulations in strong gravity in the MGR framework.

A systematic prescription to derive a PNP directly from a given spacetime metric was introduced by Mukhopadhyay \cite{Mukh02}, enabling the construction of PNPs that reproduce key general relativistic orbital properties. Pseudo-Newtonian approach has been widely used to study accretion dynamics, epicyclic frequencies, and diskoseismic modes \cite{Nowak-Wagoner, Mukho-Misra, Dihingia-SDas}. Although a single PNP cannot simultaneously reproduce both spatial and temporal properties of GR with perfect accuracy, they remain extremely valuable for physical interpretation and computational efficiency.

In the context of MGR, the importance of the PNP approach is even greater. Since the underlying spacetime geometry itself deviates from Kerr and full MGRMHD simulations are unavailable, a carefully constructed PNP provides a practical and physically consistent framework to explore accretion flows, disk structure, and particle dynamics in MGR spacetimes. Such an approach enables direct comparison between GR and MGR predictions while isolating the influence of MGR parameters.

In this work, we derive a PNP corresponding to a rotating black hole solution in a MGR theory proposed earlier \cite{ADas-Mukh}, following the metric-based prescription of Mukhopadhyay \cite{Mukh02}. We obtain the effective potential and the underlying force corresponding to PNP for both massive and massless particles in the equatorial plane, which is most relevant for accretion disk applications. The reliability of the constructed PNP is assessed by comparing various fundamental orbits and specific energy at the marginally stable orbit with their exact counterparts from the MGR spacetime.

We further investigate how the MGR parameter influences the particle dynamics and effective potentials, thereby clarifying the physical role of MGR in shaping accretion phenomena. Several fundamental spacetime properties, including deviations from Kerr geometry, are also discussed in terms of their astrophysical implications. The paper is organized as follows: Section \ref{formal} presents the derivation of the effective potential and pseudo-Newtonian force. Section \ref{test} compares the fundamental orbits obtained from MGR with those from PNP. Section \ref{imp} discusses spacetime properties and limitations of the PNP approach. A sample application of our PNP to an accretion disk around a black hole has been demonstrated in Section \ref{accretion}. Finally, Section \ref{sum} summarizes the results and outlines future prospects.

\section{Effective potential and pseudo-Newtonian force}\label{formal}

The Lagrangian density for the MGR metric (based on Eq. 32 of \cite{ADas-Mukh}) at the equatorial plane ($\theta=\pi/2$) is given by
\begin{equation}
    2L = -g_{tt}\dot{t}^2 - 2g_{t\phi}\dot{t}\dot{\phi} - g_{rr}\dot{r}^2 - g_{\phi\phi}\dot{\phi}^2.
\end{equation}
Considering the equatorial plane is justified 
because of the primary application to be in the accretion disk. We also approximate the metric upto the second order of $B/r$ and the expressions for the various components of $g_{\mu\nu}$, as we use, are given by
\begin{equation}
    g_{tt}(r) = 1 - \frac{2}{r} + \frac{3B}{r^2}-\frac{B^2}{2r^2}-\frac{33B^2}{10r^3}+O\left(\frac{B^3}{r^3}\right),
\end{equation}
\begin{equation}
    X^{\frac{1}{2}}(r) = \frac{r^2}{(\frac{B}{2}+r)^2} = 1 - \frac{B}{r}+\frac{3B^2}{4r^2}+O\left(\frac{B^3}{r^3}\right),
\end{equation}
\begin{equation}
    g_{\phi\phi}(r) = -\left[r^2 + a^{2}\left(2X^{\frac{1}{2}}(r)-g_{tt}(r)\right)\right],
\end{equation}
\begin{equation}
    g_{t\phi}(r) = a\left(X^{\frac{1}{2}}(r)-g_{tt}(r)\right),
\end{equation}
\begin{equation}
    g_{rr}(r) = \frac{-r^2}{r^2\frac{g_{tt}(r)}{X(r)} + a^2},
\end{equation}
\begin{equation}
g_{\theta\theta}(r)=r^2.
\end{equation}
The equations of motion are then
\begin{equation}
\dot{t} = \frac{(-g_{\phi\phi}E-g_{t\phi}\lambda)}{g_{t\phi}^2-g_{\phi\phi}g_{tt}},
\label{tdot}
\end{equation}
\begin{equation}
\dot{\phi} = \frac{(g_{t\phi}E+g_{tt}\lambda)}{g_{t\phi}^2-g_{\phi\phi}g_{tt}},
\label{phidot}
\end{equation}
with $E$ and $\lambda$ as the two constants of motion and the overdot implies the differentiation with respect to the proper time $\tau$.
Now from the condition $p^\mu p_\mu=-m^2$, where $p^\mu$ and $m$ are respectively 4-momentum and 
rest mass of the test particle in gravitational field, and 
by substituting $\dot{t}$ and $\dot{\phi}$ from equations (\ref{tdot}) and (\ref{phidot}), we obtain, alongside defining $R$,
\begin{eqnarray}
\nonumber
R:=\frac{\dot{r}^2}{m^2}=-\frac{1}{g_{rr}}\left(-1+g_{\phi\phi} \frac{(g_{t\phi}E_m+g_{tt}\lambda_m)^2}{(g_{t\phi}^2-g_{\phi\phi}g_{tt})^2}+2g_{t\phi} \frac{(-g_{\phi\phi}E_m-g_{t\phi}\lambda_m)(g_{t\phi}E_m
+g_{tt}\lambda_m)}{(g_{t\phi}^2-g_{\phi\phi}g_{tt})^2}\right.\\ \nonumber 
\left. +g_{tt} \frac{(-g_{\phi\phi}E_m-g_{t\phi}\lambda_m)^2}{(g_{t\phi}^2-g_{\phi\phi}g_{tt})^2}\right),\\  
\label{rdot}
\end{eqnarray}
where $E_m (=E/m)$ and $\lambda_{m}(=\lambda/m)$ are two constants of motion, specific energy and specific angular momentum, respectively. We further redefine $r/m\rightarrow r$ and define 
\begin{eqnarray}\label{massive}
    V_{eff}:=E_m^2-R,
\end{eqnarray}
which is the effective potential for a massive particle of energy $E$ in this metric. We will see some interesting consequences of the variation of $V_{eff}$ later with varying $B$ and $\lambda_{m}$.

Now, for any function $A(r)$, we define $(A)':=dA/dr$. For circular orbits,  $R=0$ and $R'=0$ (as $R=(dr/d\tau)^2$),
%(This is equivalent to $E^2=V_{eff}$ and $\frac{d}{dr}(V_{eff})=0$, where $V_{eff}=E^2-\frac{1}{2}\dot{r}^2$)%
which leads to solutions for both $E_m$ and $\lambda_m$ as functions of the orbit radius $r$, namely $E_{circ}$ and $\lambda_{circ}$ respectively, given by
\begin{equation}
E_{circ} = \frac{{g_{tt}\left(\sqrt{g_{t\phi}'^2 - g_{\phi\phi}'g_{tt}'} - g_{t\phi}'\right)+g_{t\phi}g_{tt}'}}{\sqrt{g_{tt}\left(\sqrt{g_{t\phi}'^2 - g_{\phi\phi}'g_{tt}'} - g_{t\phi}'\right)^2+2g_{t\phi}g_{tt}'\left(\sqrt{g_{t\phi}'^2 - g_{\phi\phi}'g_{tt}'} - g_{t\phi}'\right) +g_{\phi\phi}g_{tt}'^2}},
\label{em}
\end{equation}
\begin{equation}
\lambda_{circ} =\frac{-g_{\phi\phi}g_{tt}'-g_{t\phi}\left(\sqrt{g_{t\phi}'^2 - g_{\phi\phi}'g_{tt}'}-g_{t\phi}'\right)}{\sqrt{g_{tt}\left(\sqrt{g_{t\phi}'^2 - g_{\phi\phi}'g_{tt}'} - g_{t\phi}'\right)^2+2g_{t\phi}g_{tt}'\left(\sqrt{g_{t\phi}'^2 - g_{\phi\phi}'g_{tt}'} - g_{t\phi}'\right) +g_{\phi\phi}g_{tt}'^2}}.
\label{lamm}
\end{equation}
We further define the Newtonian analogue of the Keplerian angular momentum distribution $\lambda_K:=\lambda_{circ}/E_{circ}$ so that in a Keplerian accretion disk
\begin{equation}
\frac{\lambda_K^2}{r^3}=F=\frac{1}{r^3}\frac{[g_{\phi\phi}g_{tt}'+g_{t\phi}(\sqrt{g_{t\phi}'^2 - g_{\phi\phi}'g_{tt}'}-g_{t\phi}')]^2}{[{g_{tt}(\sqrt{g_{t\phi}'^2 - g_{\phi\phi}'g_{tt}'} - g_{t\phi}')+g_{t\phi}g_{tt}'}]^2}.
\label{pseudo}
\end{equation}
This $F$ is the magnitude of force corresponding to PNP in the MGR under consideration in our approach, following the prescription given in Ref.~\cite{Mukh02}.

\subsection{For a massless particle}

When $m=0$, following the above procedure, we get 
\begin{eqnarray}
\nonumber
R^0:=(\frac{dr}{d\zeta})^2=-\frac{1}{g_{rr}}\left(g_{\phi\phi} \frac{(g_{t\phi}E+g_{tt}\lambda)^2}{(g_{t\phi}^2-g_{\phi\phi}g_{tt})^2}+2g_{t\phi} \frac{(-g_{\phi\phi}E-g_{t\phi}\lambda)(g_{t\phi}E
+g_{tt}\lambda)}{(g_{t\phi}^2-g_{\phi\phi}g_{tt})^2}\right.\\ 
\left. +g_{tt} \frac{(-g_{\phi\phi}E-g_{t\phi}\lambda)^2}{(g_{t\phi}^2-g_{\phi\phi}g_{tt})^2}\right),   
\label{rdot0}
\end{eqnarray}
where $\zeta$ is an arbitrary parameter.
We also define, similarly to the massive case  (Eq. \ref{massive}), the effective potential 
\begin{equation}
    V_{eff}^0:=E_m^2-R^{0},
\end{equation}
which is the equivalent massless effective potential, with important physical implications.
From $R^0=(R^0)'=0$, we obtain 
\begin{eqnarray}
%\nonumber
\lambda_0:=\frac{\lambda_{0}^{circ}}{E_{0}^{circ}}= \frac{\sqrt{g_{t\phi}^2 - g_{\phi\phi}g_{tt}} -g_{t\phi}}{-g_{tt}},
\label{elam0}    
\end{eqnarray}
where $\lambda_{0}^{circ}$ and $E_{0}^{circ}$ are the solutions to the aformentioned condition, as obtained for the massive case.
Substituting, we obtain
\begin{eqnarray}
    F_0 = \frac{1}{r^3}(\lambda_{0})^2 = \frac{1}{r^3}\frac{(\sqrt{g_{t\phi}^2 - g_{\phi\phi}g_{tt}} -g_{t\phi})^2}{g_{tt}^2}.
\end{eqnarray}
This $F_0$ is the magnitude of force corresponding to PNP for massless particles in the MGR following the prescription given in \cite{Mukh02}. 

\section{Comparison of fundamental orbits}\label{test}

A successful PNP should reproduce the basic
properties of gravity, e.g. $r_{ms}$, $r_{mb}$, $r_{ph}$ and $E_{ms}$,
if not exactly but with small error which may be within observational uncertainties. Our PNP reproduces exactly $r_{ms}$ and $r_{ph}$ as of the exact MGR, whose values are reported in Tables \ref{tab:placeholder_label} and \ref{tab:Rph}. Our PNP also reproduces $r_{mb}$ and $E_{ms}$ within $10\%$ error, tabulated in Tables  \ref{tab:bound} and \ref{tab:ems}. Below we discuss, how are we arriving at them.
%Please refer to them as necessary.

\subsection{Marginally stable orbit}

This is an inflection point in the $V_{eff}$ profile where both
the extrema merge for a particular $\lambda_{m}$, obtained for 
$(V_{eff})'=(V_{eff})''=0$ for a circular orbit, i.e. $R=0$, ($V_{eff}=E_m^2$) under the MGR framework.

For PNP, $r_{ms}$ is obtained for $(\lambda_{K})'=0$ \cite{Mukh02}, which matches
exactly with that of exact MGR solution for all $B$ and $a$. Table \ref{tab:placeholder_label} lists the values of $r_{ms}$ for the  viable range of $a$ at various $B$.

% Requires: \usepackage{graphicx}

%\newpage

\subsection{Photon orbit}
For MGR, photon orbit radius is the inflection point for the effective potential of a massless particle, i.e. $R^{0} = 0 $ and $(R^{0})' = 0$, which can be solved to obtain the photon orbit radius under the MGR framework.

In PNP, we define the photon orbit radius as the marginally stable radius corresponding to the massless PNP, which is obtained by 
\begin{eqnarray}
 (\lambda_{0})'=0.
\end{eqnarray}

Table \ref{tab:Rph} lists $r_{ph}$ for the viable range of $a$ at various $B$. Like $r_{ms}$, all the values of $r_{ph}$ obtained from PNP match exactly with those obtained from pure MGR. 

\subsection{Marginally bound orbit}
In pure MGR, $r_{mb}$ corresponds to $E_{circ}=1$. However, for PNP, one should consider the specific energy to be zero (as the rest mass is not included therein). The specific energy in PNP at an arbitrary radius
\begin{equation}
\bar{E}=\frac{\lambda_K^2}{2r^2}+\int_{\infty}^{r} F dr.
\label{spe}
\end{equation}
The solution for $\bar{E}=0$ gives $r_{mb}$.
Table \ref{tab:bound} lists $r_{mb}$ for a viable range of $a$ at different $B$, for MGR and PNP both. While the respective values do not match, the error is less than $7\%$.

%\newpage

\subsection{Specific energy at marginally stable orbit}

We obtain $E_{ms}$ for MGR as $E_{circ}|_{r=r_{ms}}$ from Eq. (\ref{em}) and compare with that 
for PNP as $\bar{E}|_{r=r_{ms}}$ from Eq. (\ref{spe}),
which are listed in Table \ref{tab:ems}. We find that the error is within $10\%$.

\section{Important fundamental gravity properties}\label{imp}
\subsection{Analysis of correspondence between MGR and PNP}
As has been remarked earlier, in the cases of the $r_{ms}$ and $r_{ph}$, the results in MGR match exactly with those predicted by its PNP counterpart. In the cases of $r_{mb}$ and $E_{ms}$, they deviate at most by $10\%$. Therefore, the PNP is able to reproduce fundamental spatial characteristics of MGR to the point that it can be used as a replacement for full MGR in simulations of accretion and other fluid dynamical phenomena in astrophysics. 

However, PNPs tend to have a drawback in the sense that they are unable to replicate both spatial and temporal characteristics of GR at the same time. Earlier authors \cite{Mukho-Misra} showed that in GR (the $B=0$ limit of the current PNP), namely the Mukhopadhyay potential, PNP is unable to account for the radial and epicyclic frequencies of orbits predicted by the pure GR framework. Nevertheless, the introduction of $B$ as a parameter might reduce the error and, thus, it might be interesting to explore the simultaneous predictive efficiency of the PNP approach at large $B$ for temporal and spatial properties. This will be considered in a future work.

\subsection{Variation of effective potentials}
\subsubsection{Variation with B}
\begin{figure}[H]
    \centering
    \makebox[\textwidth][c]{%
        \includegraphics[width=1.2\textwidth]{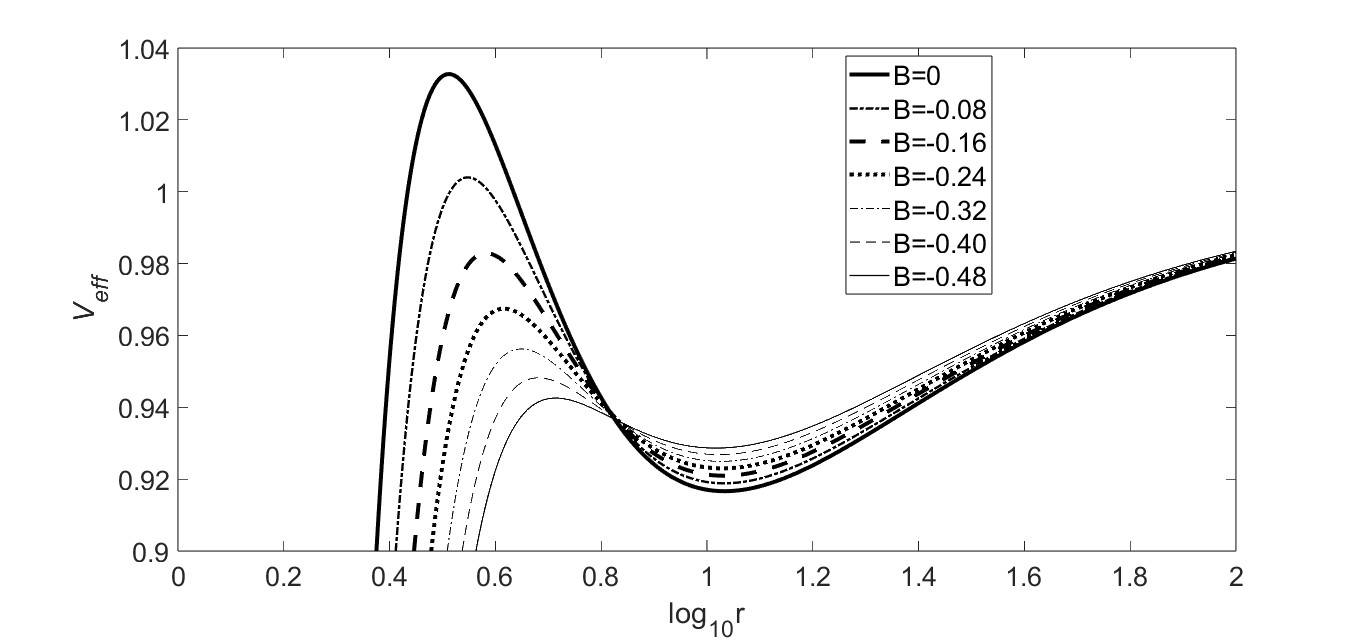}%
    }
    \caption{The variation of $V_{eff}(=E_m^2-R)$ as a function of $r$ for $a = 0.3$, $\lambda_m = 3.75$, $E_m = 1.1$ with different $B$.}
    \label{Fig1}
\end{figure}

\begin{figure}[H]
    \centering
    \makebox[\textwidth][c]{%
        \includegraphics[width=1.2\textwidth]{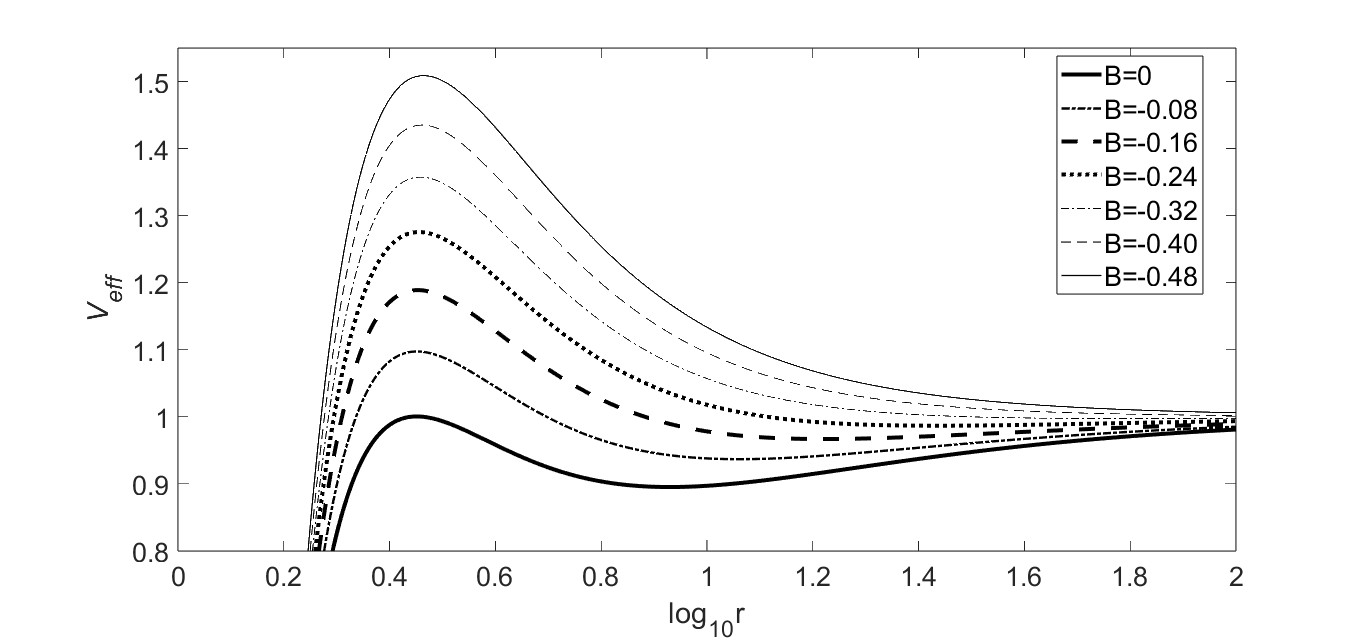}%
    }
    \caption{The variation of $V_{eff}(=E_m^2-R)$ as a function of $r$ for $a = 0.3$, $\lambda_m = 3.4$, $E_m = 1.9$ with different $B$.}
    \label{Fig2}
\end{figure}

Here we explore how the effective potentials for massive and massless particles behave and their variation with the change of parameters $B$ and $\lambda_m/\lambda_0$. 
In Figs. \ref{Fig1} and \ref{Fig2}, we show that the simultaneous increase in $E_{m}$ and decrease in $\lambda_{m}$ causes an inversion of the appearance of peak with the change of $B$. Let us try to intuit the cause of such a phenomenon, first by the $B=0$ case. Based on our description of $V_{eff}$ and by Eq. (8) from \cite{Mukh02}, we see that 
\begin{equation}\label{V_eff}
    V_{eff}|_{B=0}=\left(1-\frac{2}{r}\right)\left(1+\frac{\lambda_{m}^2}{r^2}\right)+\frac{a^2}{r^2}+\frac{4aE_{m}\lambda_{m}}{r^3}-\frac{a^2E_{m}^2}{r^3}(2+r).
    \end{equation}
Clearly, the increasing $E_{m}$ tends to decrease $V_{eff}$ irrespective of $a$. Now, let us try to see how increasing $|B|$ affects $V_{eff}$. It has been observed that for each value of $E_{m}$, there exists a value of $\lambda_{m}$ above which $V_{eff}$ resembles Fig. \ref{Fig1}. This implies that effects of $\lambda_{m}$ and $E_{m}$ work against each other by influencing different 
causes with different $|B|$. The antagonistic nature of $E_m$ and $\lambda_m$ is also apparent in Eq. (\ref{V_eff}). 

\begin{figure}[H] 
    \centering
    \makebox[\textwidth][c]{%
        \includegraphics[width=1.2\textwidth]{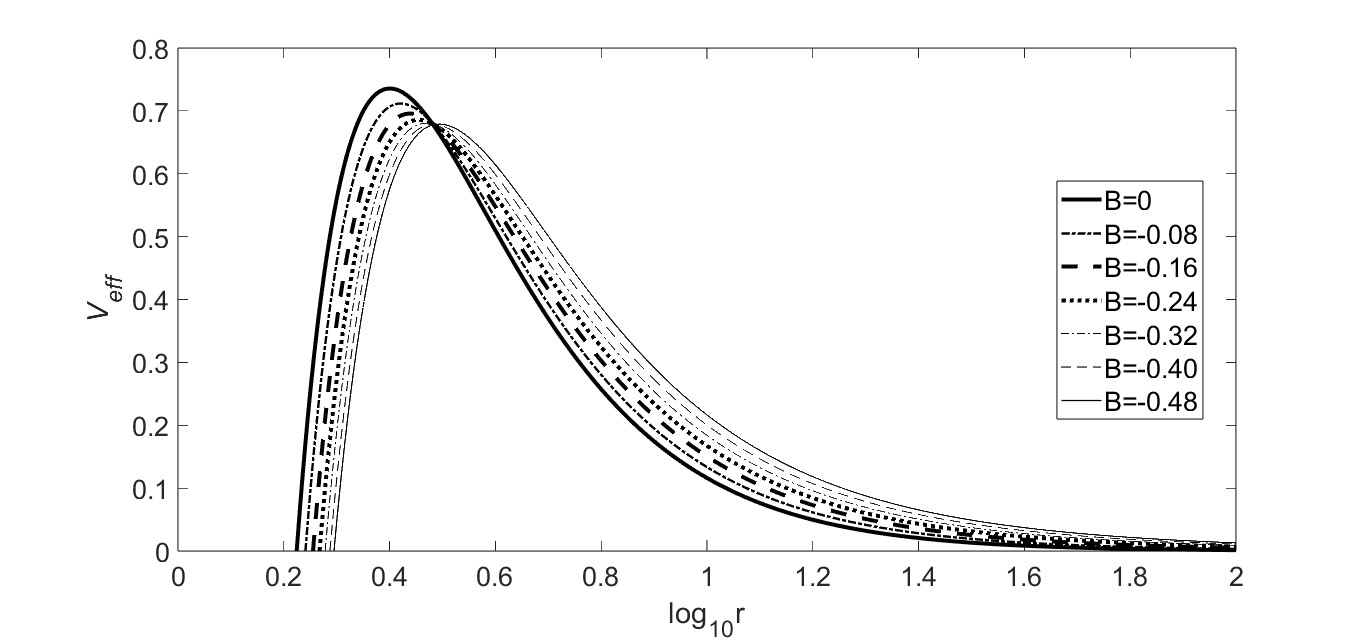}%
    }
    \caption{The same as in Fig. \ref{Fig1} except for $V^0_{eff}$.}
    \label{Fig3}
\end{figure}

\begin{figure}[H]
    \centering
    \makebox[\textwidth][c]{%
        \includegraphics[width=1.2\textwidth]{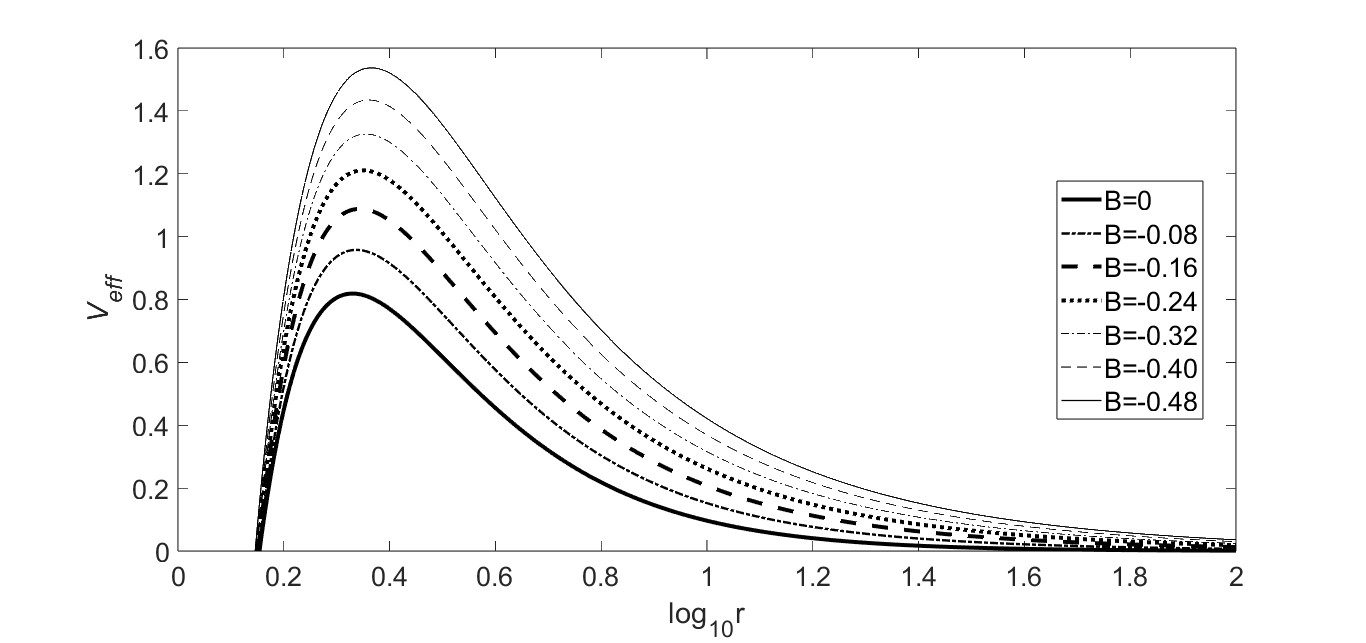}%
    }
    \caption{The same as in Fig. \ref{Fig2} except for $V^0_{eff}$.}
    \label{Fig4}
\end{figure}

Is there anything new in this antagonistic behavior for non-zero $B$? It is seen that for non-zero $B$, higher $E_{m}$ actually increases $V_{eff}$ (more precisely its peak; compare the peak of $V_{eff}$ for a lower $E_m$ with $B=0$ with the same for a higher $E_m$ with $B=-0.48$). This property can be reconciled by the hypothesis that $B$ couples to a higher power of $E_{m}$, due to its perturbative nature, than the power $E_m$ in Eq (\ref{V_eff}). This causes the given behavior. Now, if we try to get a physical notion of why this happens, it would be better to do so in the PNP picture, owing to its similarity to the Newtonian regime, which we are familiar with. It can be shown that increasing $|B|$ leads to a stronger $F$ and, thus, there is a similar antagonism between $\lambda_{m}$ and $|B|$.
See that similar nature as in Figs. \ref{Fig1} and \ref{Fig2} is seen in the massless case as well i.e. Figs. \ref{Fig3} and \ref{Fig4}. Therefore, the phenomenon is not just an artifact of mass, but of the spacetime itself.

Finally, it must be kept in mind that although PNP is able to model MGR, particularly underlying salient fluid properties, well mathematically, the complexity of MGR's physical interactions with spacetime eludes Newtonian intuition. It might be prudent to keep the physical analogy to minimum.

\subsubsection{Variation with $\lambda_{m}$}
\begin{figure}[H]
    \centering
    \makebox[\textwidth][c]{%
        \includegraphics[width=1.2\textwidth]{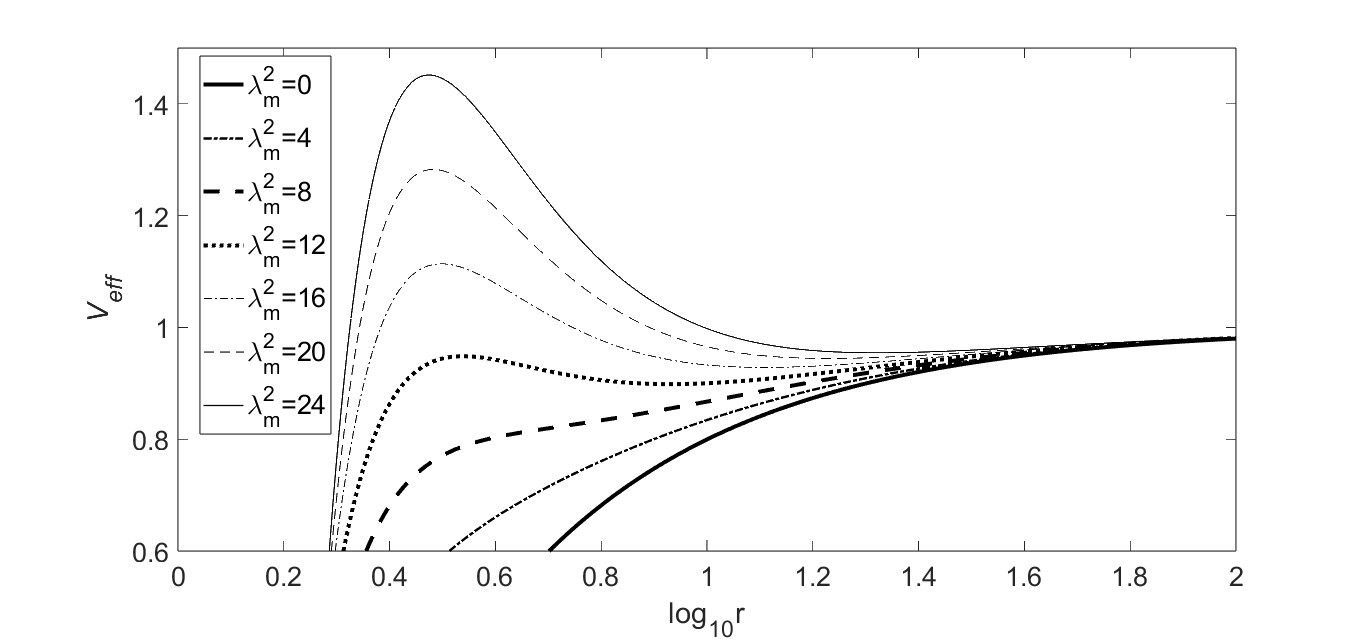}%
    }
    \caption{The variation of $V_{eff}(=E_m^2-R)$ as a function of $r$ for $a = 0.3$, $B = 0$, $E_m = 1.1$ with different $\lambda_{m}$.}
    \label{Fig5}
\end{figure}

\begin{figure}[H]
    \centering
    \makebox[\textwidth][c]{%
        \includegraphics[width=1.2\textwidth]{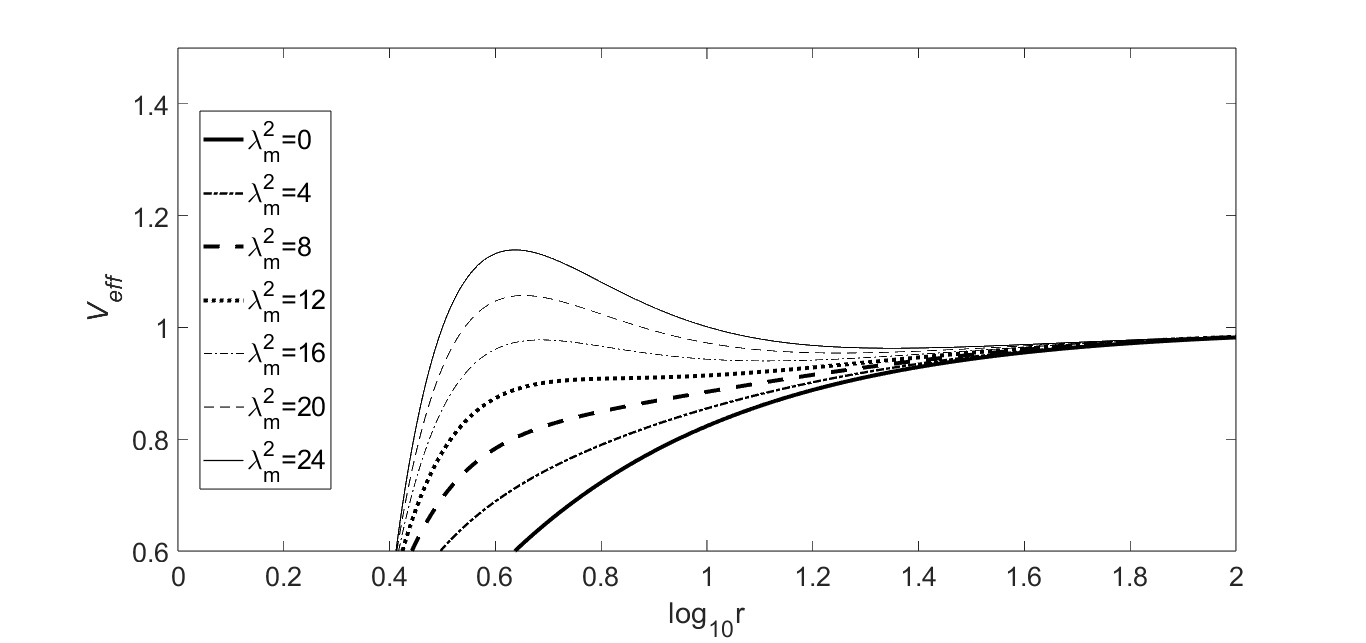}%
    }
    \caption{The same as in Fig. \ref{Fig5} except for $B=-0.5$.}
    \label{Fig6}
\end{figure}

\begin{figure}[H]
    \centering
    \makebox[\textwidth][c]{%
        \includegraphics[width=1.2\textwidth]{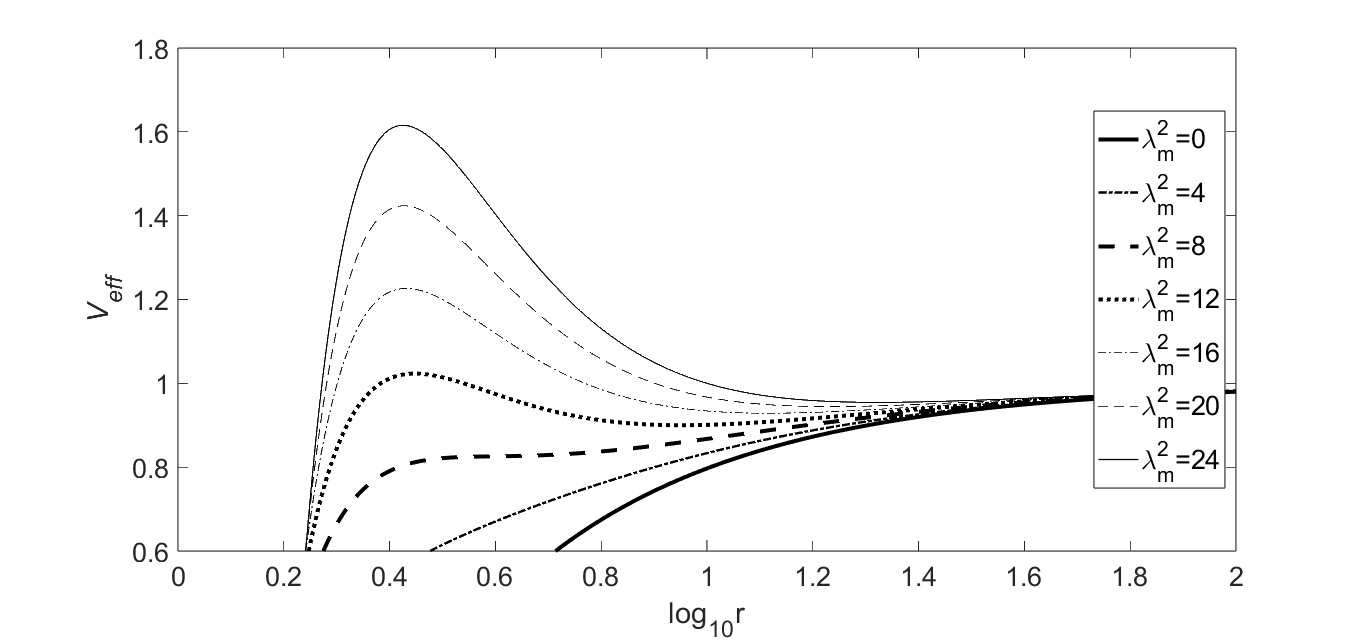}%
    }
    \caption{The variation of $V_{eff}(=E_m^2-R)$ as a function of $r$ for $a = 0.3$, $B = 0$, $E_m = 1.9$ with different $\lambda_{m}$.}
    \label{Fig7}
\end{figure}

\begin{figure}[H]
    \centering
    \makebox[\textwidth][c]{%
        \includegraphics[width=1.2\textwidth]{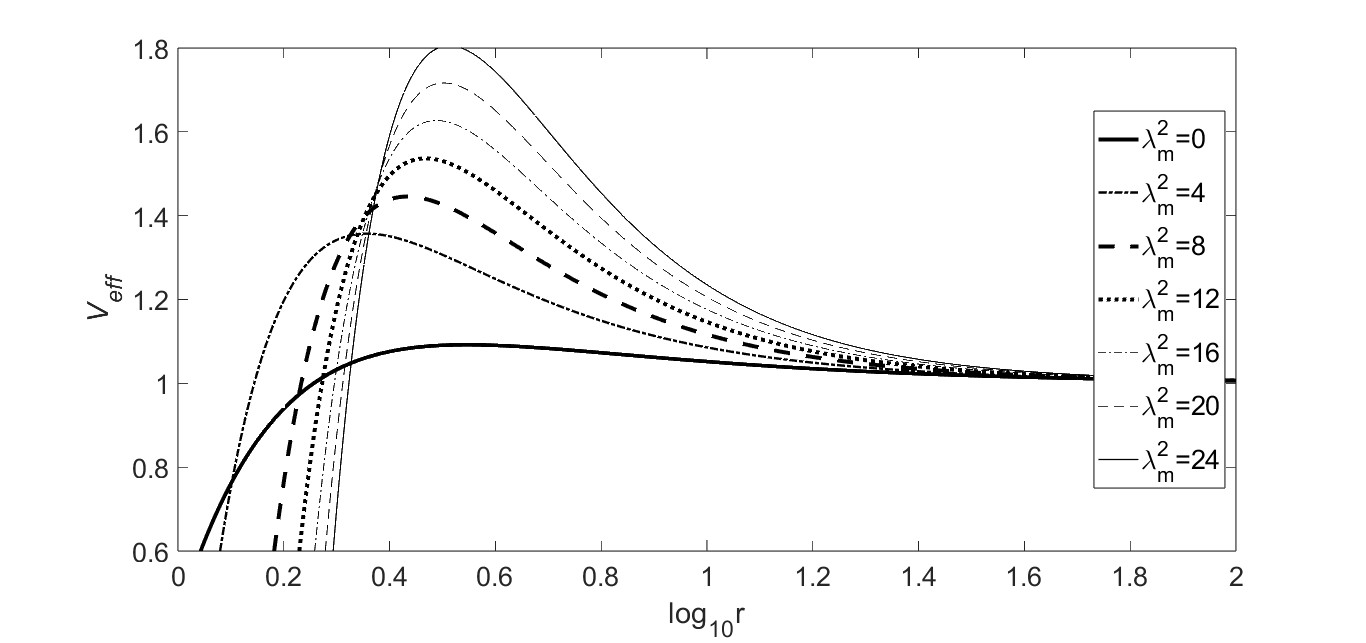}%
    }
    \caption{The same as in Fig. \ref{Fig7} except for $B=0.5$.}
    \label{Fig8}
\end{figure}

Looking at Figs. \ref{Fig5} and \ref{Fig6}, the variation of $V_{eff}$ with $\lambda_{m}$ is similar to that in GR. Increasing $\lambda_{m}$ creates a larger centrifugal barrier, which is reflected in the $V_{eff}$ and its peak. Another interesting point to note with this smaller $E_m$ is the fact that increasing $|B|$ depresses $V_{eff}$, which is another confirmation of the idea that $\lambda_{m}$ and $|B|$ play opposite roles in determining $V_{eff}$. It is as if increasing $|B|$ creates a ``spacetime pull" on the particle. However, this is true only for certain ranges of $E_{m}$ (or below a certain $E_m$). In Figs. \ref{Fig7} and \ref{Fig8}, for a higher value of $E_{m}$, we see that increasing $|B|$ has in fact increased $V_{eff}$. This is in line with the fact that $B$ couples to some higher power of $E_{m}$. While for lower values of $E_{m}$ this coupling opposes the repulsive effect in $V_{eff}$ with increasing $|B|$, at higher $E_m$ its supports repulsive effect due to the dominance of the higher order of coupling with $E_{m}$, as we can see in the corresponding figures. 

\subsection*{}
Finally, let us try to have some notion of the physics underlying the modified gravity regime. We have found that the event horizon radius for the modified black hole depends on the polar angle $\theta$, i.e. event horizon is non-rigid and non-spherical. However, it must be mentioned that the deviation from the spherical symmetry is minimal. The variation from spherical symmetry is less than $10\%$ in the event horizon radius over all possible values of $\theta$. While this is apparently different than what we know for well-known Kerr black holes, there are indeed spacetimes discussed in the literature that offer such features \cite{Ruzziconi2025, AkbarModumudi2025, DelPorro2025, Cropp2013}. There are also a few other facts such as the Killing horizon being non-trivially different from the event horizon. However, these are legitimate features of the metric obtained through the modified action approach. Despite the presence of these additional features (unconventional from the GR  perspective), our previous discussion related to PNP has been limited to $\theta=\pi/2$ where such deviations are practically even nil.

\section{Application to accretion disk}
\label{accretion}

\begin{figure}[H]
    \centering
    \makebox[\textwidth][c]{%
        \includegraphics[width=1.2\textwidth]{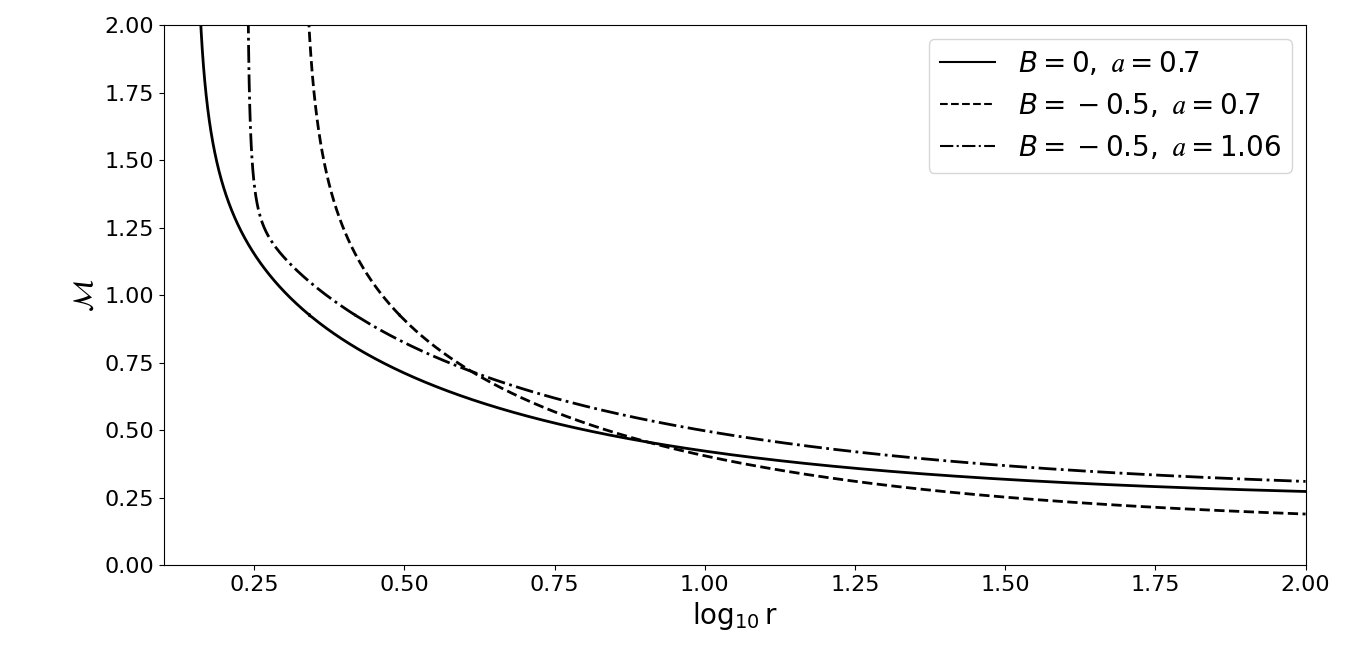}%
    }
    \caption{The variation of Mach number as a function of radial coordinate for different MGR parameter and black hole spin.}
    \label{acc}
\end{figure}

Here we demonstrate a sample application of the derived PNP for our MGR to the accretion disk.
We follow an established, well tested, model framework \cite{Mukh03}, which we do not repeat here. In short, we solve the accretion disk equations in the pseudo-Newtonian framework, but with the MGR PNP derived here. For simplicity, we further assume the flow to be inviscid and follows a polytropic equation of state \cite{Mukh03}. Fig. \ref{acc} shows how the variation of Mach number as a function of the distance from the black hole changes in the MGR with respect to that of GR. Clearly, MGR brings in the sonic transition 
away from the black hole for a given spin. This is not difficult to understand when the MGR metric leads to the larger size of the black hole. However, for a fixed MGR parameter, increasing black hole spin leads to the smaller black hole size, which in turn exhibits the sonic transition in the further inner region. This change due to the MGR effect may have many observational implications including formation of outflows/jets, particularly their launching, observed quasi-periodic oscillations in black hole sources, etc.

\section{Summary and Conclusion}
\label{sum}

In this work, a PNP corresponding to a rotating black hole solution in a MGR framework has been constructed using a metric-based prescription. The motivation stems from the practical difficulty of performing full modified-gravity GRMHD simulations, while still requiring an accurate description of general relativistic accretion dynamics. By deriving the effective potential and the underlying Newtonian-like force for both massive and massless particles, the study establishes a consistent PNP formulation suitable for accretion disk applications in MGR.

A detailed comparison between the predictions of fundamental properties of the exact MGR spacetime and those obtained from the PNP shows excellent agreement. The marginally stable orbits and photon orbits are reproduced exactly by the PNP for all considered values of the spin parameter and the modified gravity parameter. The marginally bound orbits and the specific energies at the marginally stable orbits deviate by less than about $7-10\%$, which lies within acceptable observational uncertainties. These results demonstrate that the proposed PNP successfully captures the essential spatial characteristics of the MGR spacetime, validating its use as an efficient substitute for full MGR calculations in astrophysical fluid-dynamical studies.

The behavior of the effective potential further reveals a complex interplay between the modified gravity parameter $B$, the energy and angular momentum of the particle. The analysis shows that 
$B$ can either enhance or suppress the effective potential depending on the energy regime, indicating that modified gravity introduces nontrivial corrections that cannot be interpreted through simple Newtonian intuition. Similar qualitative behavior is observed for both massive and massless particles, confirming that these effects arise from the spacetime structure itself rather than from particle properties alone. We show an example of application of our MGR based PNP to accretion disk around a black hole which exhibits a clear effect of MGR on the flow properties. 

However, although our PNP accurately reproduces the basic spatial properties, the same PNP inherits the known limitation in reproducing temporal frequencies, a point that merits future investigation, particularly for larger values of $|B|$. Alternately, another PNP for the underlying MGR has to be established in order to accurately describe temporal properties.
\section*{Acknowledgement}
The authors thank Prashant Kocherlakota of CMI and Mriganka Dutta of IISc for discussion.
SC acknowledges the
financial support from INSPIRE, DST, India. BM acknowledges a partial support by a project funded by SERB (now ANRF), India, with Ref. No. CRG/2022/003460.
\newpage

\bibliographystyle{apsrev4-2}
\bibliography{references}

%\section{Appendix}


\begin{table}[h!]
    \centering
    \begin{tabular}{|c|c|c|c|c|c|c|}
    \hline
    $a|B$   & 0       & -0.1    & -0.2    & -0.3    & -0.4    & -0.5    \\ \hline
    -1.20 & -- & -- & -- & -- & -- & 12.05\\ \hline
    -1.16 & -- & -- & -- & -- & 11.44 & 11.93 \\ \hline
    -1.12 & -- & -- & -- & 10.83 & 11.32 & 11.81 \\ \hline
    -1.08 & -- & -- & 10.22 & 10.71 & 11.20 & 11.69 \\ \hline
    -1.04 & -- & 9.61 & 10.10 & 10.59 & 11.08 & 11.57 \\ \hline
    -0.998 & 8.99 & 9.49 & 9.98 & 10.47 & 10.96 & 11.44 \\ \hline
    -0.7  & 8.14 & 8.62 & 9.10 & 9.58 & 10.06 & 10.53 \\ \hline
    -0.5  & 7.55 & 8.03 & 8.50 & 8.97 & 9.43 & 9.90 \\ \hline
    -0.3  & 6.95 & 7.41  & 7.87 & 8.34 & 8.79 & 9.25  \\ \hline
    -0.1  & 6.32 & 6.78 & 7.23 & 7.68 & 8.13 & 8.58  \\ \hline
    0     & 6.00 & 6.45 & 6.90 & 7.35 & 7.79 & 8.23  \\ \hline
    0.1   & 5.67  & 6.12 & 6.56 & 7.00 & 7.44 & 7.88  \\ \hline
    0.3   & 4.98 & 5.41 & 5.86 & 6.29 & 6.72 & 7.15  \\ \hline
    0.5   & 4.23   & 4.67 & 5.10 & 5.53 & 5.95 & 6.37  \\ \hline
    0.7   & 3.39 & 3.83 & 4.27 & 4.69 & 5.10 & 5.52  \\ \hline
    0.998 & 1.24 & 2.11 & 2.65 & 3.13 & 3.57   & 4.00  \\ \hline
    1.04 & -- & 1.57 & 2.30 & 2.83 & 3.29 & 3.73 \\ \hline
    1.08 & -- & -- & 1.81 & 2.48 & 2.07 & 3.45 \\ \hline
    1.12 & -- & -- & -- & 1.98 & 2.62 & 3.13 \\ \hline
    1.16 & -- & -- & -- & -- & 2.07 & 2.73\\ \hline
    1.20 & -- & -- & -- & -- & -- & 2.06 \\ \hline
    \end{tabular}
    \caption{$r_{ms}$ for ranges of $a$ and $B$.}
    \label{tab:placeholder_label}
\end{table}
\begin{table}[h]
    \centering
    
    \begin{tabular}{|c|c|c|c|c|c|c|}
        \hline
        $a|B$   & 0       & -0.1    & -0.2    & -0.3    & -0.4    & -0.5    \\ \hline
        -1.20 & -- & -- & -- & -- & -- & 5.17 \\ \hline
    -1.16 & -- & -- & -- & -- & 4.94 & 5.13\\ \hline
    -1.12 & -- & -- & -- & 4.71 & 4.91 & 5.10 \\ \hline
    -1.08 & -- & -- & 4.48 & 4.68 & 4.87 & 5.06 \\ \hline
    -1.04 & -- & 4.24 & 4.44 & 4.64 & 4.83 & 5.02 \\ \hline
        -0.998 & 4.00 & 4.20 & 4.40 & 4.60 & 4.79 & 4.98 \\ \hline
        -0.7 & 3.73 & 3.93 & 4.13 & 4.32 & 4.51 & 4.70 \\ \hline
        -0.5 & 3.53 & 3.73 & 3.93 & 4.12 & 4.31 & 4.50 \\ \hline
        -0.3 & 3.33 & 3.53 & 3.72 & 3.92 & 4.10 & 4.29 \\ \hline
        -0.1 & 3.11 & 3.31 & 3.51 & 3.70 & 3.89 & 4.07 \\ \hline
        0 & 3.00 & 3.20 & 3.39 & 3.58 & 3.77 & 3.95 \\ \hline
        0.1 & 2.88 & 3.08 & 3.28 & 3.47 & 3.65 & 3.84 \\ \hline
        0.3 & 2.63 & 2.83 & 3.03 & 3.22 & 3.40 & 3.58 \\ \hline
        0.5 & 2.35 & 2.55 & 2.75 & 2.94 & 3.13 & 3.31 \\\hline
        0.7 & 2.01 & 2.23 & 2.43 & 2.63 & 2.82 & 3.00 \\ \hline
        0.998 & 1.07 & 1.50 & 1.77 & 2.01 & 2.22 & 2.42 \\ 
        \hline
        1.04 & -- & 1.26 & 1.62 & 1.88 & 2.11 & 2.32 \\ \hline
    1.08 & -- & -- & 1.42 & 1.74 & 1.99 & 2.21 \\ \hline
    1.12 & -- & -- & -- & 1.53 & 1.84 & 2.08 \\ \hline
    1.16 & -- & -- & -- & -- & 1.62 & 1.92 \\ \hline
    1.20 & -- & -- & -- & -- & -- & 1.66 \\ \hline
    \end{tabular}
    \caption{$r_{ph}$ for ranges of $a$ and $B$.}
    \label{tab:Rph}
\end{table}

\begin{sidewaystable}

    \begin{center}

    \begin{tabular}{|c|c|c|c|c|c|c|c|c|c|c|c|c|}
    \hline
    \multirow{2}{*}{$a|B$} & \multicolumn{2}{c|}{0} & \multicolumn{2}{c|}{-0.1} & \multicolumn{2}{c|}{-0.2} & \multicolumn{2}{c|}{-0.3} & \multicolumn{2}{c|}{-0.4} & \multicolumn{2}{c|}{-0.5} \\ \cline{2-13} 
     & MGR & PNP & MGR & PNP & MGR & PNP & MGR & PNP & MGR & PNP & MGR & PNP \\ \hline
     -1.20 & -- & -- & -- & -- & -- & -- & -- & -- & -- & -- & 7.81 & 7.92 \\ \hline
     -1.16 & -- & -- & -- & -- & -- & -- & -- & -- & 7.41 & 7.52 & 7.73 & 7.84 \\ \hline
     -1.12 & -- & -- & -- & -- & -- & -- & 7.02 & 7.12 & 7.34 & 7.44 & 7.66 & 7.76 \\ \hline
     -1.08 & -- & -- & -- & -- & 6.62 & 6.71 & 6.94 & 7.04 & 7.27 & 7.37 & 7.59 & 7.69 \\ \hline
     -1.04 & -- & -- & 6.22 & 6.32 & 6.55 & 6.64 & 6.87 & 6.97 & 7.20 & 7.29 & 7.51571 & 7.61 \\ \hline
    -0.998 &  5.82 & 5.91 & 6.15 & 6.24 & 6.48 & 6.56 & 6.80 & 6.89 & 7.12 & 7.21 & 7.44 & 7.53 \\ \hline
    -0.7 & 5.37 & 5.39 & 5.63 & 5.69 & 5.94 & 6.01 & 6.26 & 6.32 & 6.57 & 6.63 & 6.88 & 6.94 \\ \hline
    -0.5 & 4.95 & 4.99 & 5.26 & 5.31 & 5.57 & 5.61 & 5.88 & 5.93 & 6.19 & 6.24 & 6.50 & 6.54 \\ \hline
    -0.3 & 4.58 & 4.61 & 4.89 & 4.91 & 5.19 & 5.22 & 5.50 & 5.53 & 5.80 & 5.83 & 6.10 & 6.13 \\ \hline
    - 0.1 & 4.20 & 4.21 & 4.50 & 4.51 & 4.80 & 4.81 & 5.10 & 5.11 & 5.40 & 5.41 & 5.69 & 5.70 \\ \hline
    0 & 4.00 & 4.00 & 4.30 & 4.30 & 4.60 & 4.60 & 4.89 & 4.89 & 5.19 & 5.19 & 5.48 & 5.48 \\ \hline
    0.1 & 3.80 & 3.79 & 4.09 & 4.09 & 4.39 & 4.38 & 4.68 & 4.68 & 4.97 & 4.97 & 5.26 & 5.25 \\ \hline
    0.3 & 3.37 & 3.35 & 3.67 & 3.64 & 3.96 & 3.93 & 4.25 & 4.22 & 4.53 & 4.50 & 4.81 & 4.78 \\ \hline
    0.5 & 2.91 & 2.87 & 3.21 & 3.16 & 3.49 & 3.45 & 3.78 & 3.73 & 4.06 & 4.01 & 4.33 & 4.28 \\ \hline
    0.7 & 2.40 & 2.33 & 2.69 & 2.63 & 2.98 & 2.91 & 3.26 & 3.20 & 3.54 & 3.47 & 3.81 & 3.74 \\ \hline
    0.998 & 1.09 & 1.04 & 1.63 & 1.54 & 1.98 & 1.89 & 2.30 & 2.20 & 2.59 & 2.50 & 2.87 & 2.77 \\ \hline
    1.04 & -- & -- & 1.31 & 1.23 & 1.77 & 1.68 & 2.12 & 2.02 & 2.42 & 2.32 & 2.71 & 2.61 \\ \hline
    1.08 & -- & -- & -- & -- & 1.48 & 1.39 & 1.91 & 1.80 & 2.24 & 2.13 & 2.54 & 2.43 \\ \hline
    1.12 & -- & -- & -- & -- & -- & -- & 1.61 & 1.51 & 2.02 & 1.91 & 2.34 & 2.23 \\ \hline
    1.16 & -- & -- & -- & -- & -- & -- & -- & -- & 1.70 & 1.59 & 2.11 & 1.99 \\ \hline
    1.20 & -- & -- & -- & -- & -- & -- & -- & -- & -- & -- & 1.73 & 1.63 \\ \hline
    \end{tabular}
    \caption{$r_{mb}$ for ranges of $a$ and $B$.}
    \label{tab:bound}
    \end{center}
\end{sidewaystable}

\begin{sidewaystable}

\centering
    \begin{tabular}{|c|c|c|c|c|c|c|c|c|c|c|c|c|}
    \hline
    \multirow{2}{*}{$a|B$} & \multicolumn{2}{c|}{0} & \multicolumn{2}{c|}{-0.1} & \multicolumn{2}{c|}{-0.2} & \multicolumn{2}{c|}{-0.3} & \multicolumn{2}{c|}{-0.4} & \multicolumn{2}{c|}{-0.5} \\ \cline{2-13} 
     & MGR & PNP & MGR & PNP & MGR & PNP & MGR & PNP & MGR & PNP & MGR & PNP \\ \hline
     -1.20 & -- & -- & -- & -- & -- & -- & -- & -- & -- & -- & -0.0286 & -0.0305 \\ \hline
     -1.16 & -- & -- & -- & -- & -- & -- & -- & -- & -0.0301 & -0.0322 & -0.0289 & -0.0309 \\ \hline
     -1.12 & -- & -- & -- & -- & -- & -- & -0.0317 & -0.0341 & -0.0304 & -0.0326 & -0.0292 & -0.0312 \\ \hline
     -1.08 & -- & -- & -- & -- & -0.0335 & -0.0361 & -0.0321 & -0.0344 & -0.0307 & -0.0329 & -0.0295 & -0.0316 \\ \hline
     -1.04 & -- & -- & -0.0355 & -0.0384 & -0.0339 & -0.0365 & -0.0324 & -0.0348 & -0.0311 & -0.0333 & -0.0299 & -0.0319 \\ \hline
     
     -0.998 & -0.0377 & -0.0410 & -0.0360 & -0.0389 & -0.0343 & -0.0370 & -0.0328 & -0.0353 & -0.0315 & -0.0337 & -0.0302 & -0.0323 \\
\hline
-0.7 & -0.0418 & -0.0454 & -0.0397 & -0.0429 & -0.0377 & -0.0461 & -0.0360 & -0.0387 & -0.0344 & -0.0368 & -0.0323 & -0.0352 \\
\hline
-0.5 & -0.0451 & -0.0491 & -0.0427 & -0.0462 & -0.0405 & -0.0437 & -0.0385 & -0.0414 & -0.0367 & -0.0394 & -0.0351 & -0.0375 \\
\hline
-0.3 & -0.0492 & -0.0536 & -0.0463 & -0.0502 & -0.0438 & -0.0473 & -0.0416 & -0.0447 & -0.0395 & -0.0433 & -0.0377 & -0.0403 \\
\hline
-0.1 & -0.0542 & -0.0591 & -0.0508 & -0.0552 & -0.0479 & -0.0517 & -0.0452 & -0.0487 & -0.0429 & -0.0460 & -0.0408 & -0.0436 \\
\hline
0 & -0.0572 & -0.0625 & -0.0535 & -0.0581 & -0.0503 & -0.0543 & -0.0474 & -0.0510 & -0.0449 & -0.0481 & -0.0426 & -0.0455 \\
\hline
0.1 & -0.0606 & -0.0663 & -0.0565 & -0.0615 & -0.0530 & -0.0573 & -0.0498 & -0.0536 & -0.0471 & -0.0504 & -0.0446 & -0.0476 \\
\hline
0.3 & -0.0694 & -0.0761 & -0.0641 & -0.0699 & -0.0597 & -0.065 & -0.056 & -0.0601 & -0.0524 & -0.0562 & -0.0494 & -0.0528 \\
\hline
0.5 & -0.0821 & -0.0905 & -0.0750 & -0.0819 & -0.0690 & -0.0748 & -0.0640 & -0.0689 & -0.0596 & -0.0639 & -0.0559 & -0.0596 \\
\hline
0.7 & -0.104 & -0.115 & -0.0924 & -0.101 & -0.0834 & -0.0906 & -0.0767 & -0.0822 & -0.0702 & -0.0752 & -0.0652 & -0.0694 \\
\hline
0.998 & -0.321 & -0.353 & -0.178 & -0.197 & -0.140 & -0.153 & -0.119 & -0.128 & -0.104 & -0.111 & -0.0927 & -0.0984 \\
\hline
1.04 & -- & -- & -0.251 & -0.274 & -0.165 & -0.179 & -0.133 & -0.143 & -0.114 & -0.121 & -0.100 & -0.106 \\ \hline
    1.08 & -- & -- & -- & -- & -0.218 & -0.234 & -0.154 & -0.165 & -0.127 & -0.135 & -0.110 & -0.116 \\ \hline
    1.12 & -- & -- & -- & -- & -- & -- & -0.201 & -0.212 & -0.147 & -0.156 & -0.122 & -0.128 \\ \hline
    1.16 & -- & -- & -- & -- & -- & -- & -- & -- & -0.194 & -0.201 & -0.143 & -0.149 \\ \hline
    1.20 & -- & -- & -- & -- & -- & -- & -- & -- & -- & -- & -0.199 & -0.200 \\ \hline
      \end{tabular}
    \caption{$E_{ms}$ for ranges of $a$ and $B$.}
    \label{tab:ems}

\end{sidewaystable}

\end{document}